\documentclass[twocolumn,prd,reprint,aps,groupedaddress,amsmath,amssymb,10pt,a4paper]{revtex4-2}

\usepackage{graphicx}
\usepackage{subfigure}
\usepackage{bm}
\usepackage{amssymb}
\usepackage{amsmath}
\usepackage{epsfig}

\usepackage{color}
\usepackage{mathtools}
\usepackage{cases}
\usepackage{booktabs}
\usepackage[utf8]{inputenc}
\definecolor{blue}{rgb}{0,0.3,0.8}

\usepackage[
colorlinks=true,
filecolor=black,
anchorcolor=blue,
linkcolor=blue,
citecolor=cyan, 
urlcolor=blue,
linktocpage=true,
plainpages=false,
breaklinks=true,
pdfstartview=FitH
]{hyperref}

\usepackage{bm}
\usepackage{braket}

\newcommand{\hF}{{F}}
\newcommand{\mps}{{\ell}_P}
\newcommand{\ip}{p}
\newcommand{\ii}{\hat{n}}
\newcommand{\ac}{\alpha}
\newcommand{\ab}{\varepsilon}
\newcommand{\dm}{\mathrm{d}} 
\newcommand{\RQ}{\mathcal{R_{AQ}}}

\begin{document}

\title{Superradiant instability of area-quantized Kerr black hole \\
with discrete reflectivity}
\author{Zhong-Hao Luo$^{1,3,4}$}
\email
{luozhonghao22@mails.ucas.ac.cn}
\author{Yun-Long Zhang$^{2,1,5}$}
\email
{zhangyunlong@nao.cas.cn}
\affiliation{$^{1}$School of Fundamental Physics and Mathematical Sciences,  Hangzhou   Institute for Advanced Study, UCAS, Hangzhou 310024, China.}
\affiliation{$^{2}$National Astronomy Observatories, Chinese Academy of Science, Beijing, 100101, China}
\affiliation{$^{3}$CAS Key Laboratory of Theoretical Physics, Institute of Theoretical Physics, Chinese Academy of Sciences, Beijing 100190, China.} 
\affiliation{$^{4}$University of Chinese Academy of Sciences, Beijing 100149, China.}
\affiliation{$^{5}$International Center for Theoretical Physics Asia-Pacific, Beijing/Hangzhou, China
}

\begin{abstract}
Ultralight bosons can condense to form the so-called scalar clouds around rotating black holes (BHs) through superradiant instabilities. When quantum effects near the Planck scale of the event horizon are considered, the classical BH is replaced by an exotic compact object, such as an area-quantized BH. In this work, we examine the superradiant instabilities of massive scalar fields around area-quantized BHs. We model the frequency-dependent reflectivity function of area-quantized BHs for massive scalar fields, which reflects the distinctive selection property of these BHs for massive scalar fields. We then utilize this model to investigate the case that the scalar fields can be superradiated by area-quantized BHs.  We find that the area quantization of BHs can affect the formation and effective radius, and suppress the total mass of scalar clouds. Especially, the energy gap of area-quantized BHs can break the growth continuity of scalar clouds between different modes. These are distinct from the case of classical BHs.

\end{abstract}

\date{\today}
\maketitle

\tableofcontents
\allowdisplaybreaks

\section{Introduction}
Ultralight bosons are among the most well-motivated classes of particles proposed as potential candidates for dark matter, interacting weakly with ordinary matter~\cite{Berlin:2018bsc,Arvanitaki:2009fg,Irastorza:2018dyq,Arcadi:2017kky,Schumann:2019eaa,Gaskins:2016cha}. For their extreme small masses, especially when the corresponding Compton wavelength of ultralight bosons is about (or larger than) the order of black hole (BH) scale, the ultralight bosons can be spontaneously amplified to extract the energy and angular momentum of rotating BHs to generate an exponentially growing bosonic cloud. This mechanism is known as superradiant instability~\cite{Baumann:2019eav,PhysRevD.22.2323,Starobinskii1973AmplificationOW,Brito:2015oca,Dias:2023ynv,PhysRevD.86.104017}. This bosonic cloud system, the so-called gravitational atom~\cite{Baumann:2019ztm,Baumann:2019eav,Bohra:2023vls,Arvanitaki:2009fg,Arvanitaki:2010sy}, is considered an important gravitational wave (GW) source, and it can be used to constrain the parameters of ultralight bosons through binary black hole (BBH) mergers ~\cite{Brito:2014wla,Brito:2017wnc,PhysRevD.95.043001,Tsukada:2018mbp,Zhang:2019eid,Zhu:2020tht,Sun:2020gem,Tsukada:2020lgt,Yang:2023aak,Rosa:2020uoi,Baumann:2021fkf,Tomaselli:2024dbw,Cannizzaro:2023jle}.

The renowned BH thermodynamics indicates the profound connection among gravity, quantum theory, and thermodynamics~\cite{tHooft:1999rgb,Padmanabhan:2003gd,Padmanabhan:2009vy}. The Bekenstein-Hawking entropy~\cite{Bekenstein:2008smd,PhysRevD.7.2333}, derived from BH thermodynamics, implies that the event horizon possesses more discrete degrees of freedom, which leads to the BH area quantization~\cite{Bekenstein:1995ju,Agullo:2020hxe}. Recently,
the BH area quantization has been modeled from the first principle approach in loop quantum gravity~\cite{Rovelli:1994ge,Rovelli:1996dv,Ashtekar:1997yu,Ashtekar:2000eq,Agullo:2008yv,Agullo:2010zz,FernandoBarbero:2009ai}. Moreover, area-quantized BHs have been linked to quantum modifications near the Planck scale of event horizon, commonly referred to as exotic compact objects (ECOs). In addition, various types of ECOs, such as a gravastar \cite{universe9020088}, boson star \cite{Schunck:2003kk}, and fullball \cite{Skenderis:2008qn}, have been proposed to address puzzles like the information-loss paradox~\cite{Almheiri:2012rt,Polchinski:2016hrw}.

Although the proposals are highly intriguing, it remains crucial to identify “magnifying lenses” that could bring the Planck-scale discretization of the horizon into the observable macroscopic realm. In \cite{Agullo:2020hxe},  it was shown that GW echoes and suppressed tidal heating are signs of the area quantization of BH. The GWs emitted during BH binary mergers can potentially carry information about the quantum properties of the BHs. Additionally, in \cite{Guo:2021xao,Zhou:2023sps}, it was demonstrated that the behavior of superradiance instability of scalar bosons could be affected by the
potential near-horizon new physics. Therefore, the superradiant instability can be used to detect the near-horizon microscopic structure. 

As discussed in \cite{Burgess:2018pmm,Rummel:2019ads}, the reflectivity $\mathcal{R(\omega)}$ is related to the couplings of effective field theory within the near-horizon region. In other words, the reflectivity $\mathcal{R(\omega)}$ indicates how the quantum effects affect the physics near the horizon. Therefore, we hope to investigate the impact on the superradiant instabilities of massive scalar fields by modeling the reflectivity $\RQ$ of area-quantized BHs, which can reflect the selection of quanta.

This paper is organized as follows. In Sec.~\ref{sec2}, we briefly review the discrete property of area-quantized BHs and the instabilities of massive scalar fields around ECOs. In Sec.~\ref{sec3}, we model the frequency-dependent reflectivity $\RQ$ of area-quantized BHs for scalar fields. Then, we use this model function to study the effects of area quantization on scalar clouds. In Sec.~\ref{sec4}, we conclude with a discussion and outlook of the results. In this work, we use the \((-,+,+,+)\) convention and take the unit \(G=c=1\).

\section{Area-quantized black hole and instability of massive scalar field}
\label{sec2}

Area quantization of BHs was introduced by Bekenstein in his pioneering work \cite{PhysRevD.7.2333}. Building on Christodoulou's work \cite{PhysRevLett.25.1596}, Hawking's area invariance theory \cite{PhysRevLett.26.1344}, it was recognized that the horizon area of a non-extreme BH can be considered a classical adiabatic invariant. According to Ehrenfest's principle, any classical adiabatic invariant corresponds to a quantum entity with a discrete spectrum. In Christodoulous's work \cite{PhysRevLett.25.1596}, it can be viewed as an adiabatic process when an uncharged particle crosses the event horizon of a BH at its own radial turning point. The area of this BH remains invariant, provided that the changes in other parameters are eliminated.

In \cite{PhysRevD.7.2333}, Bekenstein, invoking Heisenberg's uncertainty principle in the quantum theory, proposes that a classical point particle should be replaced by a particle with an inherent radius. Consequently, the area change of the BH is not zero but a finite value, expressed as $\Delta A_{\text{min}}=8\pi \mu b_\mu$, where \(\mu\) is the mass of this particle, and \(b_\mu\) is the inherent radius of this particle. In the classical limit \(b_\mu=0\), the classical scenario is recovered. However, a relativistic particle cannot be confined within the Compton wavelength. Therefore, the finite radius of this particle must satisfy \(b_\mu \geq \lambda_\mu\equiv \hbar/\mu\). Accordingly, the minimum area change of a BH in the quantum theory is  $\Delta A_{\text{min}}=8\pi\mps^2$, where \(\mps =\sqrt{\hbar G/c^3}\) represents the Planck scale. When the particle carries an electric charge, the minimum area change is given by $\Delta A_{\text{min}}=4\mps^2$~\cite{Hod:1998vk}. This indicates that the minimum area changes for both neutral and charged particles occur on the Planck area scale $\mps^2$. Their differing proportional coefficients in their formulas are due to the inclusion of vacuum polarization for charged particles.

\subsection{Discrete energy spectrum 
and line width}
\label{sec2a}  

Due to the quantum effect, BHs possess a uniform discrete area spectrum
\begin{align}\label{area0}
    A= {\ac} N \mps^2.
\end{align}
The phenomenological constant \({\ac}\in \mathbf{R}\) is a real number depending on concrete physical conditions. Additionally, \(A\) represents the surface area of the BH. In case of Kerr BHs, the mass \(M\), the area \(A\) and the angular momentum \(J\) obey the following relation
 \begin{align}
     M=\sqrt{\frac{A}{16\pi}+\frac{4\pi J^2}{A}}.
 \end{align}
In \cite{PhysRevD.7.2333}, Bekenstein assumes that angular momentum can also be quantized \(
J=\hbar j~(0\leq j \leq \frac{{\ac} N}{8\pi}),
\)
where \(j\) is a semi-integer. Combining it with the above area quantization in Eq.~\eqref{area0}, we can get the discrete energy spectrum of Kerr BHs

\begin{align}
    M_{N,j}=\sqrt{\hbar}\sqrt{\frac{{\ac} N}{16\pi}+\frac{4\pi j^2}{{\ac} N}}.
    \label{eq3}
\end{align}

When an area-quantized Kerr BH absorbs a quantum with the frequency \(\omega_{\ii}\) and modes \((l,m)\), it should occur to undergo the transition \(M_{N,j}\to M_{N+\Delta N,j+m}\), where $l$ and $m$ are the orbital and magnetic quantum numbers, $l\geq 0$ and $-l\leq m \leq l$. Therefore,
\begin{align}
    \hbar \omega_{\ii} 
    = M_{N+{\ii},j+m}-M_{N,j},
    \label{eq4}
\end{align}
where we set \(\Delta N={\ii}\), and \(\omega_{\ii}\) represents the ${\ii}$-th absorption spectral line.

The absorption spectrum of a macroscopic BH to scalar fields is analogously presented in \cite{Mitra:2023sny}  by comparing the case of the gravitational perturbation in \cite{Agullo:2020hxe}. In the appendix~\ref{app1}, we employ the principle of correspondence to derive the absorption spectral line $\omega_{\ii}$ in detail,
\begin{align}  
\begin{aligned}    \label{spectrum}
&\omega_{\ii}=\frac{{\ac} \kappa }{8 \pi} {\ii}+ m \Omega_H  +O(N^{-1}),\\
&\kappa=\frac{1}{2M}\frac{\sqrt{1-a^2}}{1+\sqrt{1-a^2}},
\\ 
&\Omega_H=\frac{1}{2M}\frac{a}{1+\sqrt{1-a^2}},
\end{aligned}
\end{align}
where \(\kappa\) and \(\Omega_H\) are the surface gravity and angular velocity of the Kerr BH, and the dimensionless spin parameter \(a\equiv J/M^2.\)

Additionally, the spectral line width associated with the spontaneous decay of the area-quantized BH's energy states for Hawking radiation is analogous to that observed in atomic physics \cite{Coates:2019bun}. The analytical fitting function of the line width \(\Gamma\) is presented in \cite{Datta:2021row}, where it is expressed as
\begin{align}
    \Gamma=\frac{1.005}{M} e^{-6.42+1.8a^2+1.9a^{12}-0.1a^{14}}.
    \label{eq6}
\end{align}
Thus as the BH spins up, the width of absorption spectra \(\Gamma\) also increases.

\subsection{Growth rate of superradiance around ECOs}
\label{sec2b}

Recently, the superradiant instabilities of scalar fields around ECOs have been studied in detail \cite{Guo:2021xao,Zhou:2023sps}. In this section, we employ an alternative set of fundamental solutions for ingoing and outgoing solutions with hypergeometry functions $F(-l,l+1,1-2i {\ip};z+1)$ and $F(-l-2ip,l+1-2ip,1-2i {\ip};z+1)$ given in Appendix \ref{app2}, which are different from those used in \cite{Guo:2021xao,Zhou:2023sps}. We found that the relation of the coefficients $A$ and $B$ in Eq.~\eqref{ABr} with different hypergeometry functions are still consistent with \cite{Guo:2021xao,Zhou:2023sps}. Therefore,
we obtain the same result as in \cite{Guo:2021xao}, which was given in \cite{Zhou:2023sps} after neglecting the location of the reflective membrane, as detailed in Appendix \ref{app2}. We will work in the limit \( M/\lambda_\mu=\mu M \ll 1\), where \(\mu\) and \(M\)  are the mass of the scalar field and the mass of the BH, respectively.

The metric of the Kerr BH in Boyer-Linquist coordinates \(\{t,r,\theta,\phi\}\) is
\begin{align}
\begin{aligned}\label{Kerr0}
   {\dm}s^{2} 
=
& -\left(1-\frac{2Mr}{\Sigma}\right){\dm}t^{2}+\frac{\Sigma}{\Delta}{\dm}r^{2} 
-\frac{2Mar\sin^{2}\theta}{\Sigma}{\dm}t{\dm}\phi \\& +\Sigma {\dm}\theta^{2}+\left(r^2+a_*^2+\frac{2Ma_*^2r}{\Sigma}\sin^2{\theta}\right)\sin^2{\theta} {\dm}\phi^{2},
\end{aligned}
\end{align}
where \(M\) is the BH mass,  \(
    \Sigma=r^2+a_*^2\cos^2{\theta}\), and \(
    \Delta=r^2-2Mr+a_*^2\equiv (r-r_+)(r-r_-).
\)
The quantities \(r_+\) and \(r_-\) represent the radial positions of the outer and inner horizons, respectively. Additionally, the spin parameter can defined as \(
a_*\equiv J/M =a M.
\), and the horizons are solved as $r_\pm =M[1\pm(1-a^2)^{1/2}]$

Based on the Kerr background, the massive scalar field \(\psi\) obeys the Klein-Gordon equation \begin{align}
  \nabla^\rho\nabla_\rho  \psi =\mu^2 \psi.
  \label{scalar}
\end{align}
Due to the axial symmetry inherent in the Kerr geometry, the field equation can be separable through the ansatz $\psi=e^{-i\omega t+im\phi}S(\theta)R(r).$ 
Therefore, we can obtain two separate equations for the angular function \(S(\theta)\),
\begin{align}    
\label{eq9}
&\frac{1}{\sin\theta}\frac{\dm}{{\dm}\theta}\left(\sin{\theta}\frac{{\dm}S(\theta)}{{\dm}\theta}\right)
\nonumber \\ & 
+ \biggl[ a_*^2(\omega^2-\mu^2)\cos^2{\theta} 
-\frac{m^2}{\sin^2{\theta}}+\lambda \biggr] S(\theta) =0,
\end{align}
and radial function \(R(r)\),
\begin{align}
\Delta\frac{\dm}{{\dm}r}\left(\Delta\frac{{\dm}R}{{\dm}r}\right)&+\bigg[\omega^2(r^2+a_*^2)^2-4{\omega}Mrm a_*+m^2 a_*^2  \nonumber\\ 
&-\Delta\left(\mu^2r^2+a_*^2\omega^2+\lambda\right)\bigg]R(r) =0.
    \label{eq10}
\end{align}

Here, \(\lambda\) represents the separation constant, which can be obtained as an eigenvalue from Eq.~(\ref{eq9}). Correspondingly, the eigenfunctions \(S(\theta)\) in Eq.~(\ref{eq9}) correspond to the spheroidal harmonics. It should be noted that \(\lambda\) cannot be expressed in analytical form in terms of \(l\) and \(n\), but it admits an approximate expansion given by \(\lambda\approx l(l+1)+O[a_*^2(\mu^2-\omega^2)]\). Under the conditions \(\mu M \ll 1\) and \(\omega M \ll 1\) as outlined in \cite{Starobinskii1973AmplificationOW}, the spheroidal harmonics \(S(\theta)\) reduce to the spherical harmonics \(Y^m_l(\theta)\), and the separation constant $\lambda$ is approximated as $\lambda \simeq  l(l+1)$. In particular, it is natural for the above conditions to be satisfied simultaneously. Scalar fields form bound states around the ECOs, so the kinetic energy of scalar fields is relatively small compared to the total energy. Meanwhile, based on the mass energy relationship, we have $\omega\approx\mu$. 
Therefore, if the massive scalar fields are ultralight (corresponding to the limit $\mu M\ll1$), the condition $\omega M\ll1$ is also satisfied.


The angular component \(S(\theta)\) of the complete solution given in Eq.~(\ref{scalar}) can be approximated by the well-known spherical harmonics within the limit $\mu M \ll 1$. Consequently, the crux of determining the complete solution lies in solving the radial equation, as given in Eq.~\eqref{eq10}. The radial perturbation equation can be addressed by the method of matched asymptotic expansions which involves dividing the radial component into the far region and the near region, as detailed in \cite{Cardoso:2005mh}.

In Appendix \ref{app2}, we solve the radial equation given in Eq.~\eqref{eq10} using the method of matched asymptotic expansions.
Due to the existence of the horizon, the eigenvalue of the angular frequency $\omega$ of the scalar field that forms quasi-bound states has a complex displacement $\omega=\omega_0+\delta\omega$, where $\omega_0$ represents the dominant real angular frequency of the quasi-bound states. The displacement $\delta\omega$ can enter the coefficients of the solutions about the centrifugal potential in the overlap region, as given by  
\begin{align}
\begin{split}
    R(r)=&(2kr)^{l}e^{-kr}{\hF}(-n-\delta\nu,2l+2;2kr)
    \\
 \simeq &(-1)^{n}\frac{\Gamma(2l+n+2)}{\Gamma(2l+2)}(2kr)^{l}+\\
 &(-1)^{n+1}\Gamma(n+1)\Gamma(2l+1)\delta\nu(2kr)^{-l-1},
\end{split}
\label{Rpotential}
\end{align}
where \(\nu\equiv{M\mu^2}/{k}=\nu_0+\delta\nu\) under the limit \(\delta\nu\ll1\), and \(k^2\equiv\mu^2-\omega^2\). Additionally, the form of the solutions about the centrifugal potential can convert to the form of the ingoing/outgoing solutions:
\begin{align}
\begin{aligned}
R(z)=&(Ac_1+Bc_3)U_1(0)z^{ip}\\
&+(Ac_2+Bc_4)U_2(0)z^{-ip}.
\end{aligned}
\label{Rnear}
\end{align}
Here $A,B$ are the coefficients containing $\delta\nu$ of the solutions in Eq.~\eqref{Rpotential}, $c_1,c_2,c_3,c_4$ are the transform coefficients between  different forms of the solutions.  $U_1(0)=F(-l,l+1,1-2i {\ip};1),U_2(0)=F(-l-2ip,l+1-2ip,1-2i {\ip};1)$.
Meanwhile, from the above form,
we can definite the reflectivity $\mathcal{R}$ as
\begin{align}
\mathcal{R}\equiv\frac{Ac_1+Bc_3}{Ac_2+Bc_4}\frac{U_1(0)}{U_2(0)}.
\end{align}
More details are explained in the Appendix \ref{app2}.

Finally, the eigenvalue \(\omega\) is determined as follows:
\begin{align} 
\begin{split}
&\omega =\mu \left[{1-\left(\frac{M\mu}{\nu_0+\delta\nu}\right)^2}\right]^{1/2}\simeq \omega_0+\delta\omega,\\
&\omega_0 = \mu\left[{1-\Big(\frac{\ab}{\nu_0}\Big)^2}\right]^{1/2},\\
&
\delta\omega=\frac{\delta\nu}{M}\left(\frac{\ab}{\nu_0}\right)^3 \left[1-\Big(\frac{\ab}{\nu_0}\Big)^2\right]^{1/2},
\label{angularfrequency}
\end{split}
\end{align}
where $ \nu_0\equiv l+n+1$, 
and $\ab \equiv M\mu$ represents the gravitational coupling constant.

With the reflectivity $\mathcal{R}$, we can get the growth rate of this scalar cloud 
\begin{align}
\omega_{I}= \mathrm{Im}[\delta\omega]=&\frac{\delta\nu_{*}}{M}\left(\frac{{\ab}}{\nu_0}\right)^3\left[{1-\left(\frac{\ab}{\nu_0}\right)^2}\right]^{-1/2} \nonumber\\&
\times\frac{1-|\mathcal{R}|^2}{1+|\mathcal{R}|^2+2|\mathcal{R}|\cos\phi_{\omega}},
\label{Growth rate}
\end{align}
where 
\begin{align} \label{phase1}
    \phi_{\mathcal{\omega}}
    ={i\sum_{j=1}^{l}\ln{\frac{j+2i {\ip}}{j-2i {\ip}}}} +\text{arg}\,\mathcal{R}
=2\sum_{j=1}^{l}\phi_j +\text{arg}\,\mathcal{R},
    \end{align}
with $\phi_j=\arctan\left(-\frac{2p}{j}\right)$ and ${\ip}\equiv \frac{M({ma-2\omega r_+)}}{{r_+-r_-}}=\frac{ma-2\omega r_+}{2(1-a^2)^{1/2}}$,
which agrees with \cite{Guo:2021xao} after omitting the higher order terms of $\frac{\ab}{\nu_0}$. 
Meanwhile, with $k\simeq \ab^2/( M\nu_0)$ and Eq.\eqref{eqB46}, the quantity $\delta\nu_*$ in Eq. \eqref{Growth rate} satisfies the following relation:
\begin{align}
\begin{aligned}
    \delta\nu_{*}=&2{\ip} \left[2 k\left(r_{+}-r_{-}\right)\right]^{2 l+1}\frac{(2l+n+1) !}{n!}\left[\frac{l!}{(2l) !(2l+1) !}\right]^2\\&
\times\prod_{j=1}^l\left(j^2+4 {\ip}^2\right)    \label{eq14} \\
\simeq &{(ma-2\omega r_+)} 
 \frac{(2\ab)^{4l+2}}{\nu_0^{2l+1}}
\frac{(2l+n+1) !}{n!}\left[\frac{l!}{(2l) !(2l+1) !}\right]^2 \\&
\times\prod_{j=1}^l\left[j^2(1-a^2)+  {(2\omega r_+-ma)}^2\right] .
\end{aligned}
\end{align}
 
It agrees with the result found by Detweiler in \cite{PhysRevD.22.2323}. Notice that the approximation $\lambda\simeq l(l+1)$ in Eq. \eqref{eq9} will result in an additional $1/2$ factor to $\delta\nu_*$ \cite{PhysRevD.86.104017,PhysRevD.106.064016, Zhou:2023sps}. 



Moreover, the corrected analytic results for $\delta\nu_*$ are relatively smaller compared to the numerical results \cite{PhysRevD.106.064016},
particularly at the maximum growth rate for fast-rotating BHs. Quantitative analysis demonstrates that within the $\mu M\lesssim0.35$ regime, the relative error between analytic and numerical solutions can be reduced to about $10\%$ by including the next-to-leading order correction \cite{PhysRevD.106.064016}.  Furthermore, the above approximation $\lambda\simeq l(l+1)$ is similar to the slow-rotating approximation \cite{Brito:2014wla}. Such approximation of $\lambda$ artificially suppresses higher-order contributions to the growth rate of scalar clouds, thereby reducing the capacity of scalar fields to extract rotational energy from the black hole. However, incorporating the next-to-leading order correction of $\lambda\approx l(l+1)+O[a_*^2(\mu^2-\omega^2)]$ partially reduce the relative error between analytic and numerical solutions, modestly restoring the growth rate of the scalar clouds.

In the case of ECOs, the higher order corrections affect the first part of Eq.~\eqref{Growth rate} 
\begin{align}
\frac{\delta\nu_{*}}{M}\left(\frac{{\ab}}{\nu_0}\right)^3\left[{1-\left(\frac{\ab}{\nu_0}\right)^2}\right]^{-1/2},
\end{align} 
rather than the second part
\begin{align}
  \frac{1-|\mathcal{R}|^2}{1+|\mathcal{R}|^2+2|\mathcal{R}|\cos\phi_{\omega}}.  
\end{align}
The second part is derived from the boundary condition of ECOs through the process of solving the radial equation. Meanwhile, the reflectivity $\mathcal{R}$ is assumed by the ECO model and is independent of the approximation. According to the previous discussion, the higher-order terms of $\lambda$  should enter the first part (which corresponds to classical BHs) and cause the deviation between analytical and numerical solutions. In particular, if $|\mathcal{R}|=1$, the result $\omega_{I}=0$ remains unchanged regardless of the deviations in the first part.
Furthermore, when the location of the reflective membrane near the horizon is considered at $r_0=r_{+}+\epsilon$,
the phase $\phi_\omega$ in Eq.~\eqref{phase1} is replaced by 
$    \phi_{\epsilon}
    ={i\sum_{j=1}^{l}\ln{\frac{j+2i {\ip}}{j-2i {\ip}}}}-2p\ln{\frac{\epsilon}{r_+-r_-}}+\text{arg}\,\mathcal{R}$ \cite{Zhou:2023sps}.

\subsection{Instability of massive scalar fields around ECOs}

In the context of massless scalar fields, some positive energy modes of massive scalar fields excited in the ergoregion can escape toward infinity, and the corresponding negative energy modes of the scalar fields are partially reflected by ECOs. The growth rates $M\omega_{I}\propto |\log\epsilon|^{-1}$ of massless scalar fields results in the so-called ergoregion instability \cite{Cardoso:2017njb,Maggio:2018ivz,Maggio:2017ivp}. It stems from the cascading growth of the negative energy massless scalar fields within the ergoregion of ECOs.

Therefore, it is worth considering whether there is ergoregion instability or not and what is the link for massive scalar fields between ergoregion instability and superradiant instability. The ECO possesses the reflection $|\mathcal{R}|\neq0$, the displacement $\epsilon$ in $\phi_{\epsilon}$
does not affect the sign of imaginary part $\omega_I$ in Eq. \eqref{Growth rate} of complex angular frequencies. This indicates the absence of ergoregion instability for massive scalar fields. In particular, when the ECO has a complete reflection, indicating $|\mathcal{R}|=1$, we can see $\omega_I=0$ from Eq.~\eqref{Growth rate}, which indicates the absence of the superradiant instabilities for massive scalar fields. In summary, the massive scalar fields cannot have the ergoregion instability around ECOs. Additionally, as long as the ECO does not have the complete reflection $|\mathcal{R}|=1$, there can be energy flowing outside the potential barrier to form a scalar cloud.

From a physical perspective, it is intuitive that there is no growth of scalar clouds and ergoregion instability when these massive scalar fields are reflected completely by ECOs. The positive energy modes of massive scalar fields cannot escape towards infinity due to their mass. In other words, the ECOs exert a gravitational pull that draws these positive energy massive scalar fields, which are part of the particle pairs excited within the ergoregion, back into the ergoregion. Consequently, the average vacuum fluctuation within the ergoregion is zero. Meanwhile, there is also no energy flow of massive scalar fields outside the potential barrier, preventing the growth of scalar clouds.
In Table \ref{table1}, we summarise the ergoregion and superradiant instabilities of BHs($|\mathcal{R}|=0$) and ECOs($|\mathcal{R}|=1$) .

\begin{table}[htp]
\begin{center}
\begin{tabular}{|c|c|c|c|}
\hline Scalar fields &Instability&    BHs($|\mathcal{R}|=0$) &    ECOs ($|\mathcal{R}|=1$)\\
\hline  
Massive & Ergoregion  &   Stable \cite{Brito:2015oca} &   {Stable}  \\
\hline   
Massive & Superradiant  &   Unstable\cite{Brito:2015oca} &  {Stable}\cite{Brito:2015oca}  \\
\hline   \hline   
Massless & Ergoregion   &  Stable \cite{Brito:2015oca,Maggio:2017ivp} &    Unstable \cite{Brito:2015oca,Maggio:2017ivp}  \\
\hline 
Massless & Superradiant   &  Stable \cite{Brito:2015oca} &    Stable \cite{Brito:2015oca} \\
\hline  
\end{tabular}
\end{center}
\caption{The ergoregion and superradiant instabilities of BHs($|\mathcal{R}|=0$) and ECOs($|\mathcal{R}|=1$), for massive and massless scalar fields, respectively.} 
\label{table1}
\end{table}

\section{Scalar cloud around area-quantized black holes}
\label{sec3}

In the context of area quantized BHs, it is straightforward to observe that if $|\mathcal{R}|=0$, the superradiant growth resembles that around classical BHs. Conversely, if $|\mathcal{R}|=1$, then $\mathrm{Im}[\delta\omega]=0$, indicating the absence of superradiant instabilities. Based on the above discussion, either the complete absorption indicating $|\mathcal{R}|=0$ or the complete reflection indicating $|\mathcal{R}|=1$, there is the absence of the ergoregion instability for massive scalar fields around area-quantized BHs. 

In Ref.~\cite{Datta:2021row}, the reflectivity of area-quantized BHs is described as exhibiting pulse-like behavior. The authors suggest that the reflectivity is not unity when gravitational fields cannot be absorbed by area-quantized BHs. However, this suggestion seems contradictory. If the reflectivity is to indicate the complete reflection, it should be unity outside the characteristic frequency range, rather than exhibiting pulses that suggest the partial reflection. Therefore, we will proceed to model the specific form of this reflectivity $\RQ$, and we set $\text{arg}\,(\mathcal{R_{AQ}})=0$ for simplicity.

\subsection{Reflectivity with  discrete feature} 
\label{sec3a}

\begin{figure}[h]
    \centering
\includegraphics[width=0.45\textwidth]{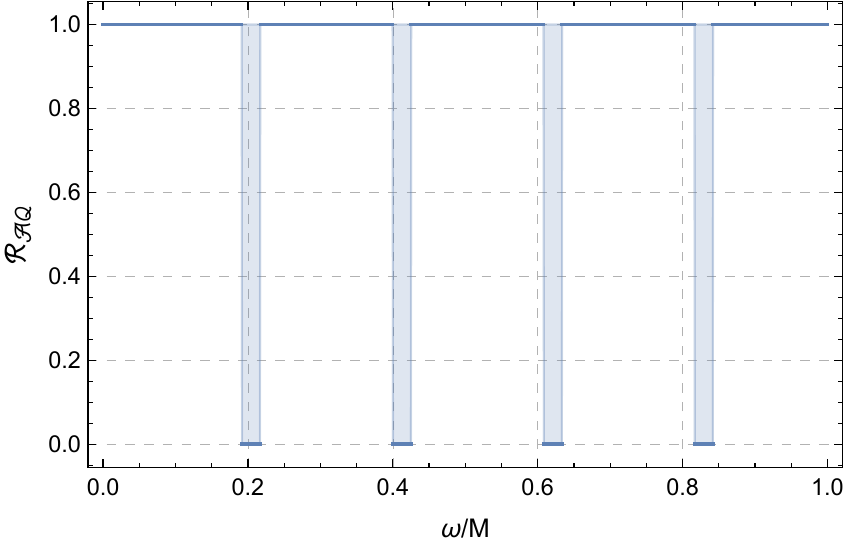}
    \caption{We plot the reflectivity of area-quantized BHs with $M=60M_{\odot}$ and $a=0.7$ for scalar fields with variable angular frequencies $\omega$, based on Eq.~\eqref{Reflectivityfunction}. The angular frequencies are normalized by the area-quantized BH mass $M$. The scalar fields with specific characterized angular frequencies can be absorbed by area-quantized BHs, resulting in $\RQ=0$. }
    \label{Romega}
\end{figure}

The reflectivity is zero within the frequency range \(f_{\ii}-\Gamma/2 < f < f_{\ii} + \Gamma/2\), and the reflectivity is unity outside the frequency range. Therefore, the reflectivity $\RQ$ can be modeled as follow:
 \begin{align}
\begin{aligned}
    &\RQ(\omega,M,a)
    \\=&
        \begin{cases}
        1-\left| \Theta\left(\sin{\zeta_{1}}\right)
        -\Theta\left(\sin{\zeta_{2}}\right) 
        \right|,&~~\Delta\omega>2\pi\Gamma,\\
\phantom{1- \text{U}\left(\sin{a}\right)}0, 
        &~~\Delta\omega\leq2\pi\Gamma,
        \end{cases}
    \end{aligned}
\label{Reflectivityfunction}
\end{align}
where
\begin{align}
    \zeta_{1}={\frac{\pi(\omega-m\Omega_{H}+2\pi\Gamma/2)}{\Delta\omega}},\\
    \zeta_{2}={\frac{\pi(\omega-m\Omega_{H}-2\pi\Gamma/2)}{\Delta\omega}},
\end{align}
with the gap of absorption spectra $\Delta\omega=\alpha\kappa/8\pi$ and $\Theta(\varphi)$ is the unit step function.

The modeled reflectivity $\RQ(\omega,M,a)$ in Eq.~\eqref{Reflectivityfunction} is characterized by three variables: the angular frequency $\omega$ of scalar fields, the mass $M$ and dimensionless spin parameter $a$ of BH. This provides two distinct perspectives for understanding the absorption of area-quantized BHs to scalar fields.

\begin{figure*}[htb]
    \centering
    \includegraphics[width=0.47\textwidth]{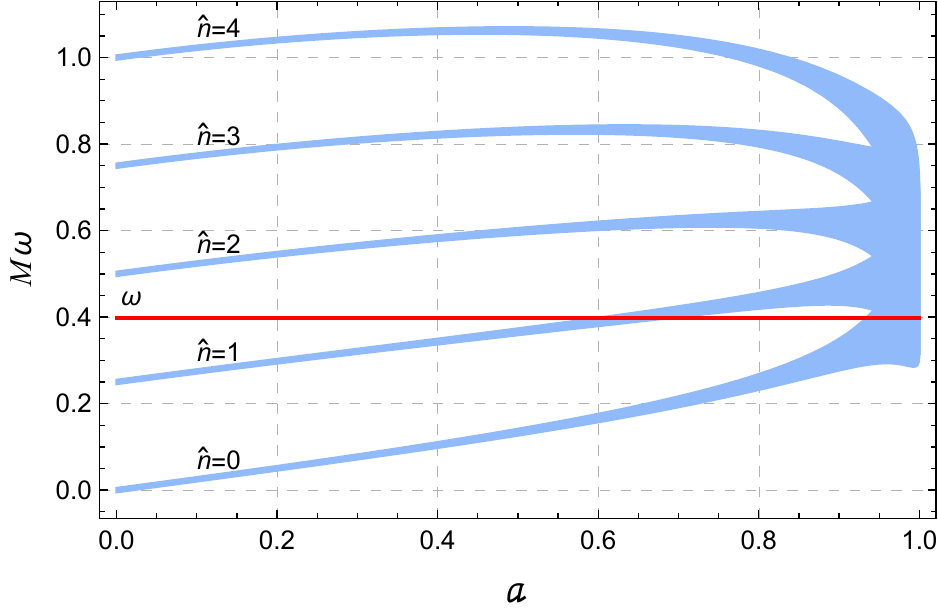}
	\quad
	\includegraphics[width=0.46\textwidth]{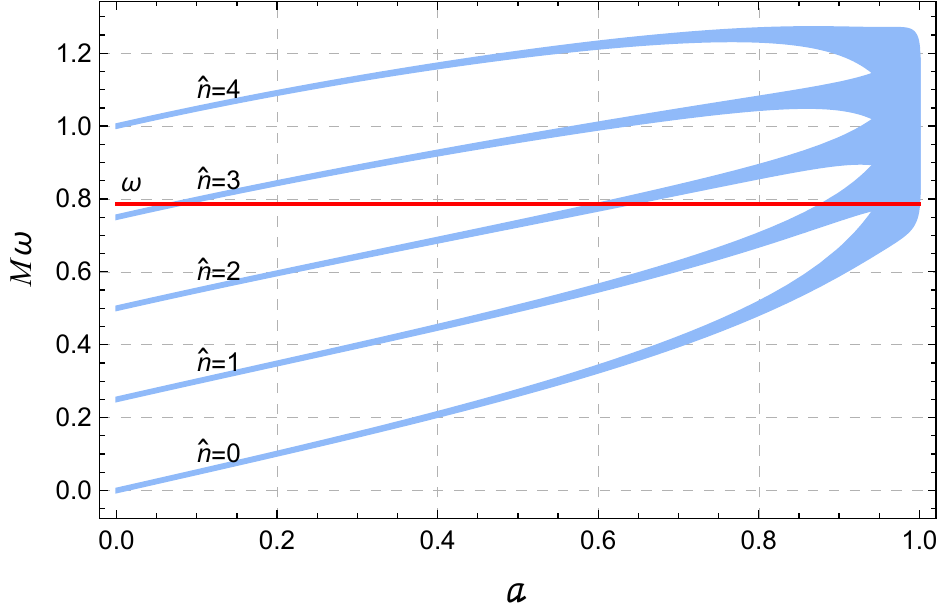}
	\includegraphics[width=0.47\textwidth]{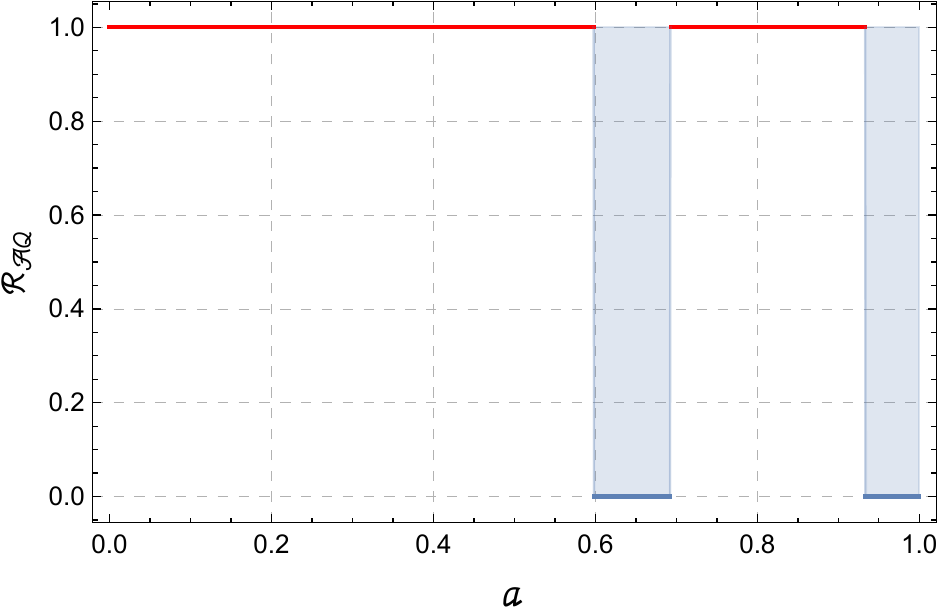}
	\quad
	\includegraphics[width=0.47\textwidth]{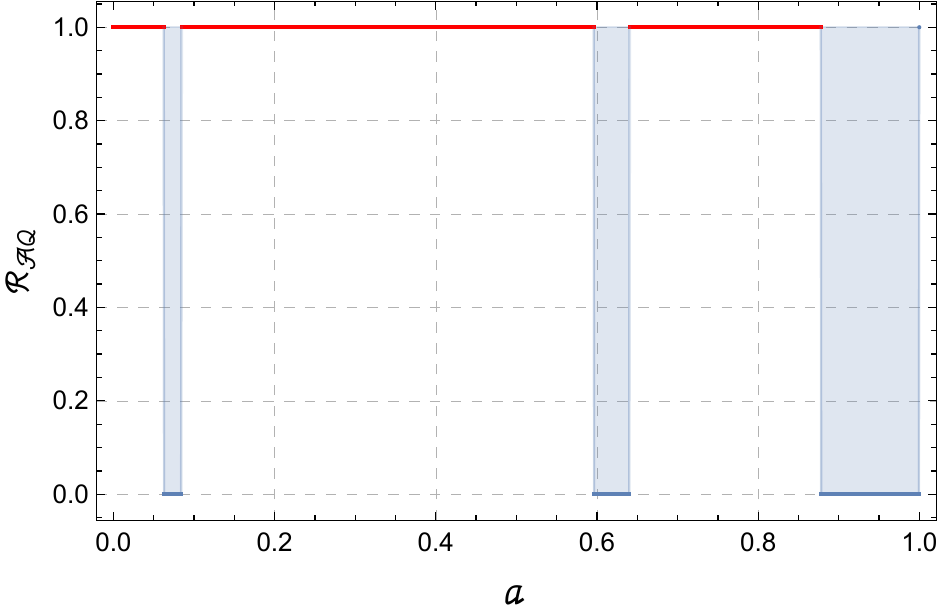}
\caption{The figures compare the reflectivities of area-quantized BHs for massive scalar fields with azimuthal quantum numbers $m=1$ (left) and $m=2$(right). In the upper figures derived from Eq.~\eqref{spectrum}, the blue region indicates the absorption spectrum of the area-quantized BHs, and the red line represents the angular frequency of the scalar fields. The lower figures (computed via Eq.~\eqref{Reflectivityfunction}) establish the reflectivity's frequency dependence: the blue lines map the resonant regimes where the scalar field's angular frequency 
spectrally overlap with BH absorption lines, 
while the red lines denote off-resonance domains. Notably, the figures contain cases beyond superradiant thresholds $\omega\geq m\Omega_H$, demonstrating that area quantization governs wave interactions independent of classical superradiance criteria.}
    \label{fig:mR}
\end{figure*}

One perspective is that: an area-quantized BH, which possesses a specific mass and spin, absorbs the scalar fields at an invariant discrete angular frequency $\omega_{\hat{n}}$ in Eq.~\eqref{spectrum}, due to the fixed value of the angular velocity $\Omega_H(a,M)$ and the surface gravity $\kappa(a,M)$. 
This reflects the selection of scalar fields by an area-quantized BH. The absorption of the scalar fields with $\omega_{\hat{n}}$ leads to the energy transitions of the area-quantized BH and the reflectivity of the BH for the scalar fields $\RQ=0$.  
This process is illustrated in Fig.~\ref{Romega}.

The other perspective is that: the scalar fields with a specific energy $\hbar\omega$ interacts with area-quantized BHs with the mass $M$. The discrete angular frequencies $\omega_{\hat{n}}$ vary with the spin $a$ of area-quantized BHs. This perspective is similar to the study of gravitational-wave echoes for area-quantized BHs in \cite{Agullo:2020hxe}. As in \cite{Agullo:2020hxe}, only if the oscillation frequency in Eq.~(7) of the quasi-normal modes (QNMs) matches the energy lines of area-quantized BHs, the absorption of the QNMs (echoes are expected) occurs for values of the rotation parameter $a$ between the consecutive intersection of the red line and black lines in their Fig.~1. The difference is that, in Fig.~1 of \cite{Agullo:2020hxe}, the red curve indicates gravitational QNMs as a function of $a$, 
in our Fig.~\ref{fig:mR},
the horizontal red line indicates a constant angular frequency of scalar fields.

Subsequently, it is important to check if the modeled reflectivity $\RQ$ conforms to the absorption of scalar fields by area-quantized BHs as plotted in the top row of Fig.~\ref{fig:mR}. We can see that the intersection regions in the upper diagrams of Fig.~\ref{fig:mR}, corresponding to the absorption of the scalar fields with a specific energy by the area-quantized BH, align with the region of $\RQ=0$ in the bottom diagrams of Fig.~\ref{fig:mR}.
Therefore, to determine the full spin range of the BHs that can absorb the scalar fields with specific energy, one does not need to plot all the absorption curves as depicted in the upper diagrams of Fig.~\ref{fig:mR} to analyze the number and position of intersection points. Instead, one can simply set the
reflectivity $\RQ(\omega,M,a)=0$ in Eq.\eqref{Reflectivityfunction},
input the angular frequency $\omega$ of the scalar fields and the BH mass $M$, and solve for the entire range of BH dimensionless spin parameter $a$.

It is noteworthy that the absorption spectrum maybe overlap as the rotation of the BHs accelerates, leading to an increase in line width beyond the absorption angular frequency gap $\Delta\omega=\alpha\kappa/8\pi$. Our calculation reveal that the absorption spectrum overlaps only when the  dimensionless spin parameter \(a_{\text{crit}}\geq0.9430\). In such case,  area-quantized BHs can recover to classical BHs.

\subsection{The limited effective radius of the scalar cloud}
\label{sec3b}
\begin{figure}[h]
    \centering
\includegraphics[width=0.45\textwidth]{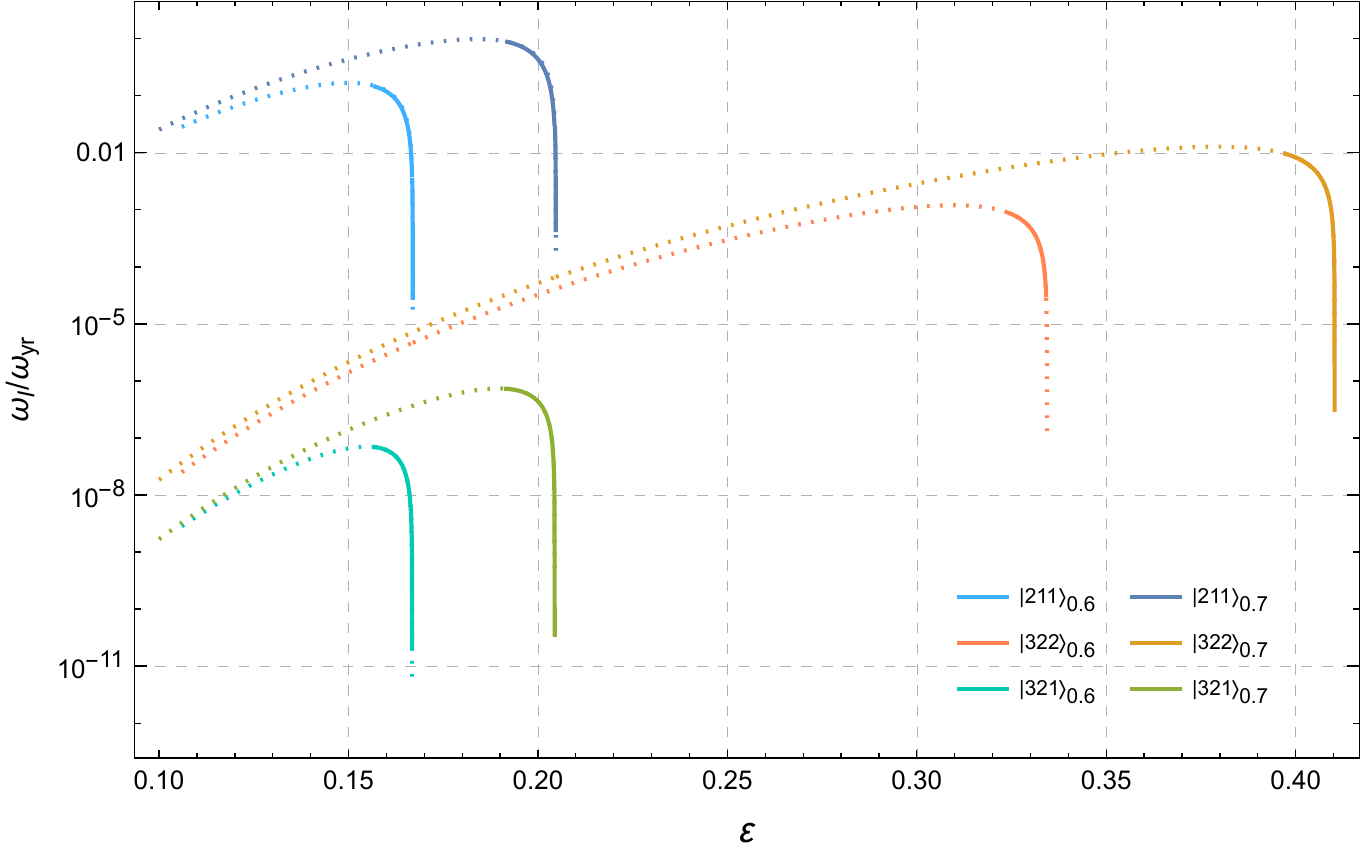}
\caption{We plot the superradiance growth rates \(\omega_{I}/\omega_{\text{yr}}\) as the function of $\varepsilon$, where
    \(\omega_{I}\equiv \mathrm{Im}[\delta\omega]\) in Eq.~\eqref{Growth rate} with $\RQ=\RQ(\omega(\varepsilon),M,a)$. It is normalised by $\omega_{\text{yr}}\equiv 2\pi/T_{\text{yr}}$ with $T_{\text{yr}}$ representing the period of one year. The dotted curves illustrate the growth rates of scalar clouds around classical BHs, whereas the solid curves depict those for area-quantized BHs with a mass of $60M_{\odot}$. With two BH dimensionless spin parameters \(a=0.6\) and \(a=0.7\), the growth rates of different three fundamental modes $\ket{nlm}$ depend on the gravitational coupling constant \({\ab}\). It is noteworthy that the value of $\omega_I/\omega_{\text{yr}}$ should be smaller than the corresponding numerical result near their maximum \cite{PhysRevD.106.064016}.}
    \label{SP1}
\end{figure}

In this section, from the above first perspective, we demonstrate the impact of the selection of scalar fields stemming from the invariant BH energy gap on the effective radius of scalar clouds. By substituting the frequency-dependent $\RQ(\omega(\varepsilon),M,a)$ with the specific BH mass $M$ and the dimensionless parameter $a$ into the superradiant growth rate $\omega_{I}$ as given in Eq.~(\ref{Growth rate}), we analyze the variation of $\omega_{I}$ with respect to the gravitational coupling constant $\varepsilon$. The results are illustrated in Fig.~\ref{SP1}. In the following text, we use the notation $\ket{nlm}$ to label the modes of scalar clouds with the real part of angular frequency $\omega_0$ in Eq.~\eqref{angularfrequency}.

In Fig.~\ref{SP1}, we observe that the regions where $\omega_{I}>0$ are bounded by two endpoints. The right endpoint at $\varepsilon_{0}$ corresponds to the superradiance condition $\omega(\varepsilon_0)=m\Omega_{H}$,
marking the critical point where the $\omega_{I}$ transitions from positive to negative. This transition is governed by the parameter $p$ in Eq.~(\ref{eq14}), which dictates the sign of superradiant growth rates, and its relation is presented in \cite{Jia:2023see}, as follows:
\begin{align}
p &=\frac{M({ma-2\omega r_+)}}{{r_+-r_-}}=\frac{2Mr_{+}}{r_{+}-r_{-}}(m\Omega_{H}-\omega)\nonumber \\&
=-\frac{\omega}{16\pi}\frac{\delta A}{\delta M}.
    \label{pA}
\end{align}
This critical point is also present in the context of classical BHs. Notably, the relation only holds when the thermal effects of BHs are not considered. Eq.~\eqref{pA} connects the superradiant instabilities with Hawking's BH area theorem which suggests that without the thermal effects of BHs, the total horizon area of a BH cannot decrease over time: $\delta A\geq0$ \cite{PhysRevLett.26.1344}. The formation of scalar clouds with the superradiant condition $\omega<m\Omega_{H}$, leading to the growth rates $\omega_I>0$ (or $p>0$), can extract the energy of BHs. From Eq.~\eqref{pA}, this implies the BH area theorem $\delta A>0$ with $\delta M<0$. 

However, in the context of area-quantized BHs, the superradiant condition cannot hold because the energy transitions of area-quantized BHs only occur at this condition $\omega=m\Omega_{H}$. This condition results in the zero growth rates of scalar clouds $\omega_I=0$ (or $p=0$) and the zero area change $\delta A=0$. This indicates that if there are no Hawking thermal effects for area-quantized BHs, there will be no growth of scalar clouds around them. Meanwhile, the surface area of area-quantized BHs will not change.  

Therefore, the growth of scalar clouds around area-quantized BHs is linked with the Hawking radiation of area-quantized BHs. The left endpoint at $\varepsilon_{\Gamma}$ corresponds to $\omega(\varepsilon_\Gamma)=m\Omega_{H}-\pi\Gamma$, determined by the line width $\Gamma$ from Hawking radiation. At this point, the parameter $p$ is a function of the line width $\Gamma$:
\begin{align}
    p(\Gamma)=\frac{2Mr_{+}}{r_{+}-r_{-}}\pi\Gamma=\frac{\pi\Gamma}{2\kappa}.
\end{align}
Thus, from Eqs.~\eqref{Growth rate}-\eqref{eq14} we can see that the growth rate of scalar clouds $\omega_{I}(\Gamma)$ is also a function of the line width $\Gamma$. 
Next, we  calculate $\varepsilon_{0}$ where $\omega_{I}=0$, which is determined by the condition $\omega(\varepsilon_0)=m\Omega_{H}$. 
With Eq.\eqref{spectrum}  we obtain \begin{align}
\varepsilon_{0}\simeq \frac{ma}{2(1+\sqrt{1-a^2})}. 
\end{align} Subsequently, the shift $\Delta\varepsilon\equiv \varepsilon_{0}-\varepsilon_{\Gamma}$ is expressed as
\begin{align}
    \Delta\varepsilon\approx M\pi\Gamma=1.005\pi e^{-6.42+1.8a^2+1.9a^{12}-0.1a^{14}}.
\end{align}

For the usual dimensionless spin parameters $a$, the shift $\Delta\varepsilon$ is minor. For instance, at $a=0.7$, $\varepsilon_{0}=0.204$ (with $m=1$) and  $\Delta\varepsilon=0.013$, which is the order of $\sim\mathcal{O}(0.01)$ . Consequently, $\varepsilon$ is confined to a single value within a broad range. Moreover, the value of $\omega_{I}$  at $\varepsilon_{\Gamma}$ is close to the peak value shown in Fig.~\ref{SP1}, and the precise determination of the peak's extreme point is computationally complex. Therefore,  $\varepsilon_{\Gamma}$ can be used to represent the confined range of $\varepsilon$.

In summary, the energy gap of area-quantized BHs leads to the confined region of gravitational coupling constant $\varepsilon$, and then the constrained value of $\varepsilon$ results in the scalar clouds around area-quantized BHs having a definite effective radius. We employ the Bohr radius,
\begin{align}
    r_{c}\equiv \frac{\lambda_\mu}{\varepsilon_{\Gamma}}=\frac{\hbar M}{{\varepsilon_{\Gamma}}^2}=\frac{\hbar M}{(ma/2(1+\sqrt{1-a^2})-M\pi\Gamma)^2},
\end{align}
to describe the characteristic spatial separation of the scalar cloud from the area-quantized BH. 
It is evident that the modes with the same $m$ have the same radius, which is approximately $1/m^2$ of the radius for the $m=1$ mode. In addition, as the dimensionless spin parameter $a$ increases, the radius $r_{c}$ of the scalar clouds decreases.

\subsection{Suppression and total mass of scalar cloud}
\label{sec3c}

\begin{figure*}[t]
    \centering
    \includegraphics[width=0.45\textwidth]{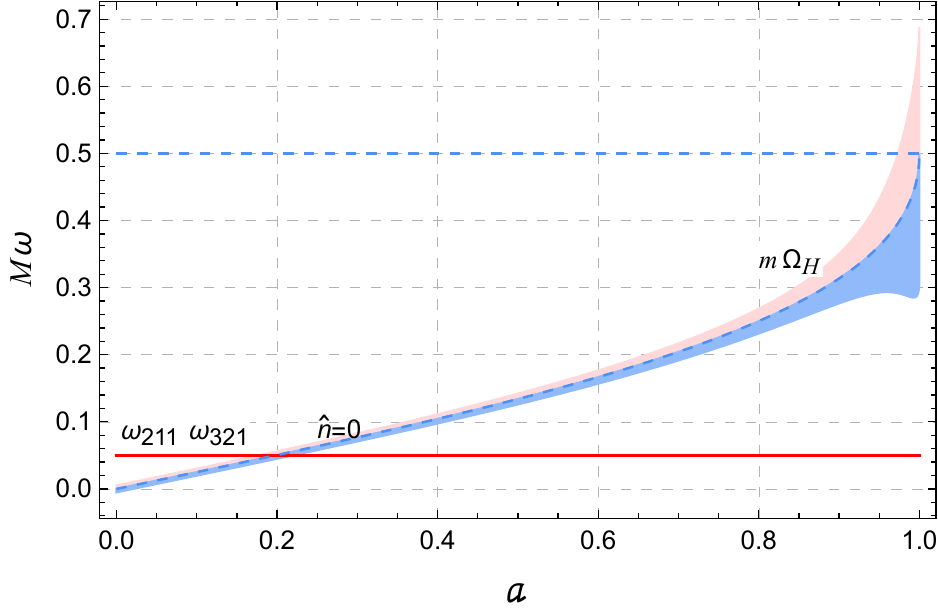}
    \includegraphics[width=0.45\textwidth]{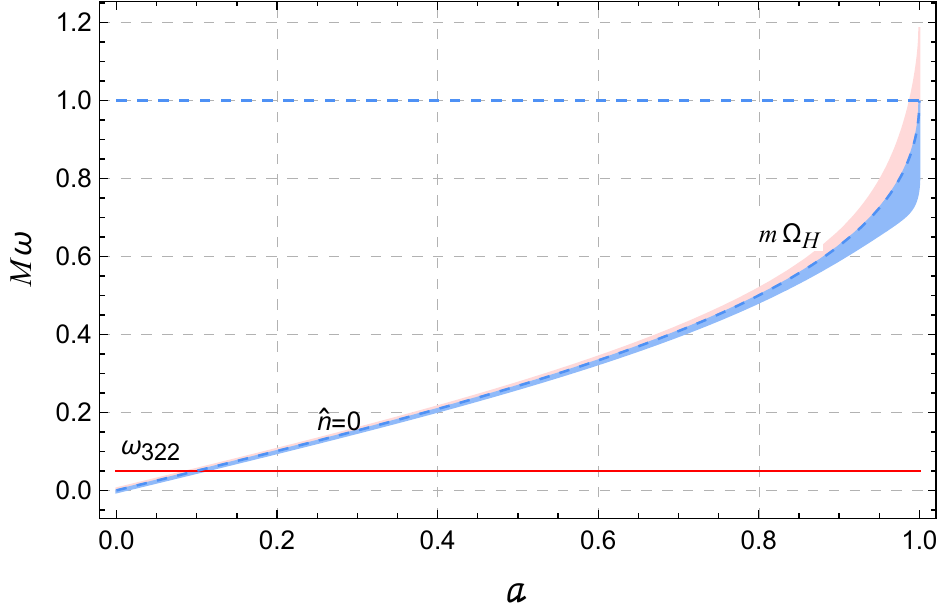}
    \includegraphics[width=0.65\textwidth]{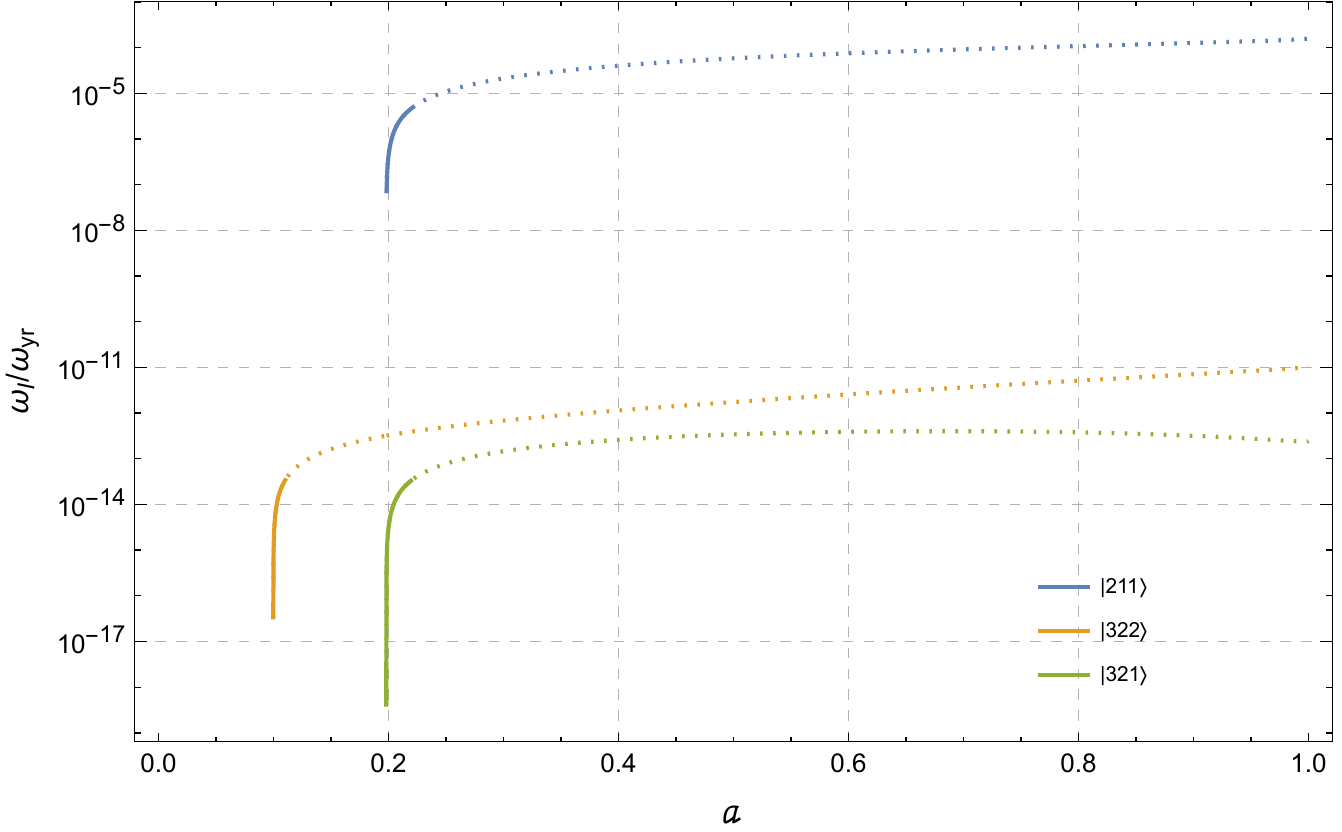}
\caption{The top figures visualize the first absorption spectral line from Eq.~\eqref{spectrum}, where the red band ($m\Omega_H<\omega<m\Omega_H+\pi\Gamma$) and the blue band ($ m\Omega_H-\pi\Gamma<\omega<m\Omega_H$)  denote the upper and lower broadening regimes of the spectral line, respectively. 
In the bottom
figure, we quantify the superradiant growth rates $\omega_{I}/\omega_{\text{yr}}$ via Eq.~\eqref{Growth rate}.
For BHs with \(M=60M_{\odot}\), we plot the  growth rates of the fundamental modes \(\ket{211}\), \(\ket{322}\) and \(\ket{321}\) as a function of the spin $a$, optimized through $\omega_0\approx\varepsilon M=0.05M$ to maximize analytic-numerical consistency($\Delta|\omega_{I}/\omega_{\text{yr}}|<3\%$).
Crucially, the consistency between the consecutive intersections($\RQ=0$) in the upper figures and the solid curves in the bottom figure reveals area quantization signatures – with solid curves representing area-quantized BHs and dashed curves classical counterparts. 
}
    \label{SP2}
\end{figure*}


In the previous subsection, we discuss the superradiant instabilities of scalar fields around the area-quantized BHs with a specific mass $\mu$ and spin $a$, which have the invariant BH energy gap. However, if we consider the second perspective mentioned by Sec.~\ref{sec3a}, which suggests the interaction of the scalar fields with a specific energy $\hbar\omega$ towards area-quantized BHs with a mass $M$. In this context, the spin of area-quantized BHs can change the energy gap of the BH transitions from Eq.~\eqref{spectrum} and Eq.~\eqref{eq6}, which is equivalent to the transition lines $\omega_{\hat{n}}(a)$ being a function of the dimensionless spin parameter $a$.

From Sec.~\ref{sec3b}, we have observed that there are superradiant instabilities for the scalar fields around area-quantized BHs, only if their angular frequencies satisfy the following condition:
\begin{align}
    \omega_{\hat{n}=0}(a)-\pi\Gamma(a)<\omega<\omega_{\hat{n}=0}(a),
    \label{eq23}
\end{align}
where $\omega_{\hat{n}=0}=m\Omega_H$ is the first energy transition line of area-quantized BHs for scalar fields. Thus, the scalar fields exhibit the superradiant instabilities around the area-quantized BHs only within a narrow range of the dimensionless spin parameter $a$, as shown in Fig.~\ref{SP1}. 

The restricted range of the dimensionless spin parameter $a$ can suppress the formation of the scalar cloud outside this range.
For instance, as observed in Fig.~\ref{SP2} for the BHs with a slightly larger dimensionless spin parameter, such as $a=0.4$, a scalar cloud would be expected to form in the case of classical BHs for $\omega_{I}>0$. However, no scalar cloud forms around area-quantized BHs at this value of $a$, as shown in the bottom figure for $\omega_{I}=0$.

Additionally, the scalar clouds forming around area-quantized BHs are expected to be lighter due to the suppression outside the confined spin range. Here, we estimate the total mass of the scalar clouds around area-quantized BHs. Since the rate of gravitational wave radiation from the scalar cloud is generally slower than the growth rate of the scalar cloud, we can neglect the gravitational wave radiation. Under this approximation, the mass and angular momentum of both the BH and the scalar cloud are conserved, as expressed by
\begin{align}
    M_{cloud}=M_{i}-M_{f},\quad J_{cloud}=J_{i}-J_{f},
    \label{eq32}
\end{align}
where $M_{cloud}$ and $J_{cloud}$ represent the mass and angular momentum of the scalar cloud, and the subscripts $i$ and $f$ denote the initial and final states of the BH, respectively.

{}
At infinity, the scalar field consists of many quanta, each with the energy $E=\hbar\omega$ and the angular momentum $J=\hbar m$. Thus, the ratio of the angular momentum and the energy carried by the scalar field is $m/\omega$, which can be expressed as
\begin{align}
    \frac{\Delta J}{\Delta M}=\frac{m}{\omega}.
\end{align}
Substitute this into Eq.~(\ref{eq32}), we obtain \begin{align}
    \frac{m}{\omega}(M_{i}-M_{f})=J_{i}-J_{f}.
    \label{eq34}
\end{align}
At this point, the scalar cloud ceases to grow, and the superradiant instabilities become saturated when 
$\omega=m\Omega_{H}=\frac{ma_f}{2M_f(1+\sqrt{1-a_{f}^2})}$,
which can be solved as $a_f=4 m \omega M_f /(m^2+4\omega^2 M_f^2 )$.
Substituting it with $J_i=a_i M_i^2$  and $J_f=a_f M_f^2$ into Eq.~(\ref{eq34}), we obtain the final mass of the BH
\begin{align}
    M_{f}=\frac{m^3-\sqrt{m^6-16m^2\omega^{2}M_{i}^{2}(m-\omega M_{i}a_{i})^2}}{8\omega^{2}M_{i}(m-\omega M_{i}a_{i})},
\end{align}
where $a_{i}$ is the initial dimensionless spin parameter of the BH. Given $\ab\ll1$, the first-order approximate expansion for the final mass is
\begin{align}
    M_{f}\approx M_{i}-\frac{ \omega M_i^2}{m}\left( a_i-\frac{4\omega M_i}{m} \right)+O(\ab^3).
\end{align}
Thus, the total mass of the scalar cloud is approximately
\begin{align}
    \frac{M_{cloud}}{M_{i}}\simeq\frac{\mu M_i }{m} \left( a_i-\frac{4\mu M_i}{m} \right)+O(\ab^3).
    \label{eq38}
\end{align}

From the above result, the total mass of the scalar cloud $M_{cloud}$ depends on the initial BH spin $a_i$, once the superradiant instabilities terminate.
The initial dimensionless spin parameter $a_i$ can be computed using the lower bound of the angular frequency $m\Omega_{H}(a_{i})-\pi\Gamma(a_{i})=\omega\approx\mu$.
This result in \eqref{eq38} agrees with the derivations in \cite{Branco:2023frw},
where
$M_{cloud}/M\simeq ({a}_{i}-{a}_{f})\mu M/m$, $a_f\simeq 4\mu M/m$ and the approximate invarient mass has been used.
Besides, we can employ the more efficient method described in Sec.~\ref{sec2} to determine the initial dimensionless spin parameter $a_{i}$ by setting 
$\RQ(\omega, M, a)=0$ and solving for the value of the dimensionless spin parameter $a$ yields the maximum value, which corresponds to the initial dimensionless spin parameter $a_i$.

In addition to strictly numerically solving the initial dimensionless spin parameter $a_i$, we can roughly estimate the order of magnitude of the ratio of the scalar cloud mass to the mass of area-quantized BH $M_{cloud}/M_i$. From Eq.~\eqref{eq23}, we can get the range of the area-quantized BH spin that can trigger the superradiant instabilities:
\begin{align}
    \frac{2\mu r_+(a_f)}{m}\leq a\leq \frac{2[\mu+\pi\Gamma(a_i)]r_+(a_i)}{m}
\end{align}
Due to $a_f\simeq a_i$ in the context of area-quantized BHs, we can get $a_i-a_f\simeq 2\pi\Gamma(a_i) r_+(a_i)/m$. Thus, the ratio of the scalar cloud mass to the mass of area-quantized BH is roughly
\begin{align}\label{eq31}
    \frac{M_{cloud}}{M_i}\simeq \frac{2\pi\Gamma(a_i) r_+(a_i)\mu M_i}{m^2}\sim \frac{\mu M_i}{m} \times O(10^{-2}),
\end{align}
where $\mu M_i/m$ is the bound in the context of classical BHs. Therefore, the mass of the scalar clouds around the area-quantized BHs is approximately two orders of magnitude lighter than the case of classical BHs.

\subsection{Disrupted growth continuity of scalar clouds between different modes}
\label{sec3d}

{}

As scalar clouds form around BHs continuously, they extract the energy of the BHs. The scalar clouds, with different modes, can grow around BHs one after another until the BHs stop spinning. However, in the context of area-quantized BHs, the continuity may be broken by the energy gap of area-quantized BHs. Generally, the mass of area-quantized BHs extracted by scalar clouds in \eqref{eq31}, which reaches about $0.1\%$ of the BH mass, is approximately two orders of magnitude lighter than the case of classical BHs. For simplicity, we focus on the variation of the spin of area-quantized BHs neglecting the influence of the mass change of the BHs. 

Initially, an area-quantized BH, with a dimensionless spin parameter $a_i$, possesses the energy gap $\hbar(m\Omega_{H}(a_{i})-\pi\Gamma(a_{i}))=\Delta E_{min}\leq\Delta E\leq \Delta E_{max}=\hbar m\Omega_H(a_i)$. When the energy of the scalar fields $\hbar\omega_0=\hbar\mu\sqrt{1-\left(\frac{\ab}{l+n+1}\right)^2}$ is almost equal to this energy gap $\Delta E$ of the BH, the scalar fields can form a scalar clouds around the BH. Particularly, if the energy of the scalar fields just satisfies the minimal energy gap of the BH, namely $\hbar\omega_0=\Delta E_{min}$, the initial dimensionless parameter $a_i$ is the maximum dimensionless spin parameter where the scalar clouds can start to form at this moment. Subsequently, the formation of scalar clouds makes the area-quantized BH spin down. The spin deceleration of the BH will reduce the transition energy gap $\Delta E$ of the BH. When the dimensionless spin parameter of the BH reduces to $a_f$, the maximum energy gap $\Delta E_{max}=\hbar m\Omega_H(a_f)=\hbar\omega_0$ reach the found of superradiant condition $\omega_0<m\Omega_{H}$. This indicates that the superradiant instabilities have saturated. 

The above initial and final dimensionless spin parameters $a_i$ and $a_f$ can be computed numerically by using the following equations:
\begin{align}
\begin{split}
m\Omega_{H}(a_{i})-\pi\Gamma(a_{i})&=\omega_0,\\
m\Omega_{H}(a_f)&=\omega_0.
\end{split}
\end{align}
Based on the relations above, we can observe if a scalar cloud with the $\ket{211}$ mode has formed around an area-quantized BH with a dimensionless spin parameter $a_{f\ket{211}}$, where $\hbar\Omega_{H}(a_{f\ket{211}})=\omega_0$. The minimal energy gaps $\Delta E_{322}(a)\equiv \hbar (2 \Omega_H(a)-\pi \Gamma(a))$ at $a_{f\ket{211}}$ and  $a_{i\ket{322}}$ of the area-quantized BH are
\begin{align}
\begin{aligned}
\Delta E_{322}(a_{f\ket{211}})&= 2 \hbar\Omega_H(a_{f\ket{211}})-\pi\hbar\Gamma(a_{f\ket{211}})
\\
&=2\hbar\omega_0  -\pi\hbar \Gamma(a_{f\ket{211}}),
\\
\Delta E_{322}(a_{f\ket{322}})&=2\hbar\Omega_H(a_{f\ket{322}})-\pi \hbar\Gamma(a_{f\ket{322}})\\
&=\hbar\omega_0,
\end{aligned}
\end{align}
 where $a_{i\ket{322}}$ is the initial dimensionless spin parameter, which makes the scalar clouds with $\ket{322}$ mode start to grow.
However, 
in the discrete region of the area-quantized BHs, $\pi\Gamma(a_{f\ket{211}})\ll \omega_0$, which lead to $\Delta E_{322}(a_{f\ket{211}})>\Delta E_{322}(a_{i\ket{322}})$.
This is also depicted in Figure \ref{SP2}, that $a_{f\ket{211}}$ cannot translate to $a_{i\ket{322}}$ immediately if the area-quantized BH is isolate. Thus, the superradiant instabilities of the $\ket{322}$ mode cannot be triggered even if the scalar clouds of the $\ket{211}$ mode become saturated.

To illustrate this process intuitively, we provide an example for an area-quantized BH with a mass of $M=60M_{\odot}$. Initially, for the  $\ket{211}$ mode, which is the most rapidly growing,  the scalar clouds can only form around the area-quantized BH with the initial dimensionless spin parameter \(a_{i}=0.220\). As the BH spins down, the growth rate $\omega_{I}$ of the $\ket{211}$ mode approaches zero when the dimensionless spin parameter of the BH decreases to \(a_{f}=0.198\), where $\omega_{211}=m\Omega_{H}$. Meanwhile, the total mass of the scalar cloud with the $\ket{211}$ mode is given by $M_{cloud}/M_i=0.11\%$.
Subsequently, If the BHs were classical, the scalar cloud with the $\ket{322}$ mode would be expected to form next. However, the scalar cloud with the $\ket{322}$ mode cannot form at \(a=0.198\) where $\omega_{I}=0$. 

Thus, we observe a significant and distinct phenomenon considering the suppressed formation of scalar clouds around area-quantized BHs. Specifically,
once the superradiant growth of the $\ket{211}$ mode reaches saturated, the formation of scalar clouds with other modes around the isolated area-quantized BHs is inhibited. This is markedly different from the behavior around isolated classical BHs, where the scalar clouds with other modes sequentially form, one after another.

\section{Conclusion}\label{sec4}
Based on the proposal by Bekenstein and Mukhanov~\cite{Bekenstein:1995ju}, area-quantized BH exhibits non-zero and discrete reflectivity for scalar fields. The non-zero reflectivity influences the absorption of scalar fields by area-quantized BHs. If the scalar fields, which consist of these scalar fields, can also be superradiated by the BHs, the area quantization of BHs can significantly affect the superradiant instabilities of the scalar fields around area-quantized BHs. Specifically, this results in the confined radius, suppressed formation, reduced mass and disrupted growth continuity of the scalar clouds. 

In this work, we have supplemented the derivation related to the important results, specifically the expressions in Eq.~(\ref{spectrum}) and Eq.~(\ref{Growth rate}), as presented in Sec.~\ref{sec2}. These supplements, detailed in appendix \ref{app1} and \ref{app2}, render these results more convincing. Moreover, we discuss the connection of the two instabilities of massive scalar fields, the ergoregion and superradiant instabilities,  around ECOs. Subsequently, we model the reflectivity $\RQ$ of area-quantized Kerr BHs for massive scalar fields. We consider this modeled $\RQ$ from two distinct perspectives. From the first perspective, an area-quantized BH absorbs the scalar fields within an invariant angular frequency $\omega_{\hat{n}}$ leading to a transition. From this viewpoint, we observe the confined effective radius of scalar clouds around area-quantized BHs. 

Meanwhile, we reveal the significant link between the superradiant instabilities and Hawking thermal effects in the context of area-quantized BHs. The second perspective involves the interaction of the scalar fields carrying a specific energy $\hbar\omega$ with an area-quantized BH with a mass of $M$. In this context, the spin of area-quantized BHs can change the energy gap of the BH transitions. Consequently, this can result in the suppressed formation and reduced radius, as well as disrupted growth continuity of the scalar clouds, which are significantly distinct from the case of classical BHs.

A point worthy of further discussion is the influence of the astronomical environment when an area-quantized BH is no longer isolated. For example, there is an accretion disk or more matter providing spin for the BH. In addition, the little continuous variation of the area-quantized BH mass cannot be considered in this research. Moreover, the scalar clouds surrounding the BHs can radiate gravitational waves or even the fast radio bursts~\cite{Rosa:2017ury,Chen:2023bne}, which provides interesting gravitational phenomenology of astronomical observations \cite{Guo:2023mel, Cao:2023fyv}. 
It will also be interesting to consider the vector or tensor condensates
\cite{Cardoso:2019mes,Jia:2023see}.
In the following work, we hope to study these effects of area quantization to effectively probe those astrophysical BH systems.

\begin{acknowledgments}
This work is supported by the National Key Research and Development Program of China (No. 2023YFC2206200), the National Natural Science Foundation of China (No.12375059), and the Fundamental Research Funds for the Central Universities (No.E2ET0209X2). We are grateful to the anonymous referee for many valuable suggestions, which helped a lot to improve the  manuscript.
\end{acknowledgments}

\bigskip

\appendix

\section{The absorption
approximate formula}\label{app1}

Here we derive the absorption approximate formula in Eq.~(\ref{eq4}) under a large $N$ limit.
Approximately, Eq.~(\ref{eq4}) can be expanded as
\begin{align}
&M_{N+\Delta N, j+\Delta j}=\sqrt{\hbar} \sqrt{\frac{{\ac}(N+\Delta N)}{16 \pi}+\frac{4 \pi(j+\Delta j)^2}{{\ac} (N+\Delta N) }} \nonumber\\
& \simeq  M_{N, j}+\frac{\partial M_{N, j}}{\partial N} \Delta N+\frac{\partial M_{N, j}}{\partial j} \Delta j+O\bigg(\frac{\Delta N}{N}, \frac{\Delta j}{j}\bigg).
\end{align}
Substituting the above equation into Eq.~(\ref{eq4}), we get 
\begin{align}
&M_{N+\Delta N, j+\Delta j}-M_{N, j} 
\simeq   \frac{\partial M_{N, j}}{\partial N} \Delta N+\frac{\partial M_{N, j}}{\partial j} \Delta j\nonumber \\
&=  \frac{1}{2M}\left(\frac{ \hbar{\ac}}{16 \pi}-\frac{4 \pi \hbar j^2}{ {\ac} N^2}\right) \Delta N+\frac{1}{2M}\left(\frac{8 \pi \hbar j}{{\ac} N }\right) \Delta j.
\label{eqA2}
\end{align}
Consider the following relations
\begin{align}
\begin{aligned}
&A=8\pi M^2(1+\sqrt{1-a^2}),\\
&j=\frac{aM^2}{\hbar}, ~(0\leq j \leq \frac{{\ac} N}{8\pi}),\\
&N=A/{ \hbar{\ac}},
\end{aligned}
\end{align}
the first term in Eq.~(\ref{eqA2}) becomes
\begin{align}
\begin{aligned}
&\frac{1}{2 M}\left(\frac{ \hbar{\ac}}{16 \pi}-\frac{4 \pi \hbar j^2}{{\ac} N^2 }\right) 
=  \frac{ \hbar{\ac}}{16\pi M} \frac{\sqrt{1-a^2}}{1+\sqrt{1-a^2}}
=\frac{\hbar{\ac} \kappa }{8 \pi} ,
\end{aligned}
\end{align}
where \(\kappa\equiv\frac{1}{2M}\frac{\sqrt{1-a^2}}{1+\sqrt{1-a^2}}\) is the surface gravitation of the Kerr BH. The second term in Eq.~(\ref{eqA2}) becomes 
\begin{align}
\begin{aligned}
    &\frac{1}{2 M}\left(\frac{8 \pi \hbar j}{{{\ac}}N}\right) 
=\frac{\hbar a}{2 M\left(1+\sqrt{1-a^2}\right)}  =\hbar \Omega_H,
\end{aligned}
\end{align}
where \(\Omega_H\equiv\frac{1}{2M}\frac{a}{1+\sqrt{1-a^2}}\) is the surface angular velocity of the Kerr BH. 

So with \( \Delta N={\ii}, \Delta j=m\),  we can get 
\begin{align}
    \begin{aligned}\label{homega}
\hbar \omega_{\ii}
&\simeq \frac{\partial M_{N, j}}{\partial N} \Delta N+\frac{\partial M_{N, j}}{\partial j} \Delta j+O\bigg(\frac{\Delta N}{N}, \frac{\Delta j}{j}\bigg) \\
& \simeq \hbar{\ii} \frac{{\ac} \kappa}{8 \pi} \, +\hbar m \Omega_H +O\left(N^{-1}\right).
\end{aligned}
\end{align}
Furthermore, due to the correspondence principle, the approximate expansion also indeed satisfies the macroscopic first law of Kerr BH
\begin{align}
    \mathrm{d} M=\frac{\kappa }{8 \pi} \mathrm{d} A +\Omega_{H} \mathrm{d} J=\hbar{\ii} \frac{\kappa {\ac}}{8 \pi} + \hbar m \Omega_H,
\end{align}
which is consistent with Eq.~(\ref{homega}) considering $\mathrm{d} M=\hbar  \omega_{\ii}$.

\section{Solving the radial equation}
\label{app2}

In this appendix, we solve the radial equation as given in Eq.~(\ref{eq10}) with the matched asymptotic expansion method, which includes the far region, the near region, and the overlap region solutions.

\subsection{The far region solution}

In the far region, where the influence from the BH can be ignored, the equation can be approximated to a Schrödinger-like equation with a Newtonian potential. In the far region limit, \(r\gg M\) and \(r \sim l/\omega\), the terms  in the radial equation can be approximated as
\begin{align}
\begin{aligned}
&\frac{\Delta\omega^2a_{*}^2}{r^4}\sim 0,\quad
\frac{{\omega}Mm a_{*}}{r^3}\sim 0,\quad
\frac{m^2a_{*}^2}{r^4}\sim0,
   \\ 
&\frac{\omega^2(a_{*}^2+r^2)^2}{r^4} \simeq \omega^2, \quad\frac{\Delta\mu^2r^2}{r^4} \simeq \mu^2\bigg(1-\frac{2M}{r}\bigg),\\ &\frac{\Delta\frac{d}{dr}\left(\Delta\frac{dR}{dr}\right)}{r^4} \simeq  \frac{d^2R}{dr^2}+\frac{2}{r}\frac{dR}{dr}, 
\frac{\Delta l(l+1)}{r^4} \simeq  \frac{l(l+1)}{r^2}.
\end{aligned}
\end{align}
Then the radial equation in  Eq.~\eqref{eq10} can be approximated to the radial hydrogen atom like equation
\begin{align}
    \frac{d^2(rR)}{dr^2}+\left[\omega^2-\mu^2\big(1-\frac{2M}{r}\big)-\frac{l(l+1)}{r^2}\right](rR)=0.
\end{align}
After setting \(\rho=2kr\), we obtain
\begin{align}\label{rhoR}
    \frac{d^2(\rho R)}{d\rho^2}+\left[-\frac{1}{4}+\frac{\nu}{\rho}-\frac{l(l+1)}{\rho^2}\right](\rho R)=0,
\end{align}
where the notations
\begin{align}
k\equiv\sqrt{\mu^2-\omega^2},\quad
\nu\equiv{M\mu^2}/{k}.
\end{align}
The solution of Eq.~\eqref{rhoR} is
\begin{align}\rho R(\rho)=e^{-\rho/2}\rho^{l+1}F(l+1-\nu,2l+2;\rho),
\end{align}
namely
\begin{align}
    R(r)=(2kr)^l e^{-kr} F(l+1-\nu,2l+2;2kr),
    \label{eqB7}
\end{align}
where \(F(l+1-\nu,2l+2; 2kr)\) is the confluent hypergeometric function.

Unlike the boundary condition of the electron wave function in the hydrogen atom, which is located at the origin, the inner boundary condition of this problem is located at the event horizon, which corresponds to absorption by the BH. So the eigenvalues \(\omega\) should be the complex formula and have corresponding complex \(\nu\) \cite{Berti:2009kk}, which satisfy
\begin{align}
    \nu=    \nu_0+  \delta\nu,\quad   \nu_0\equiv n+ l+1,
    \label{eqB8}
\end{align}
where \(\delta\nu\) is a small complex correction.
Then, the solution in Eq.~(\ref{eqB7}) also can be expressed as 
\begin{align}
    R(r)=(2kr)^l e^{-kr}F(-n-\delta \nu,2l+2;2kr).
\end{align}

\subsection{The near region solution}
\label{ab2}

In the near region, namely the region close to the event horizon, the solution contains the information about the event horizon of the BH. Because \(r=r_+\) is the singularity of the radial equation, as \(r\) approach to the event horizon, the function \(R\) in radial equation Eq.~(\ref{eq10}) varies very rapidly. It is convenient to introduce the rescaled radial coordinate to accommodate this rapid change
\begin{align}
    z\equiv \frac{r-r_+}{r_+-r_-},\qquad
    \Delta=z(z+1)(r_+-r_-)^2.
\end{align}
When \(r\!\ll\!l/\omega\) or \(r\!\ll\!l/\mu\), namely \(z\!\ll\!l/\omega(r_+-r_-)\) or \(z\!\ll\!l/\mu(r_+-r_-)\), then
\begin{align}
    \begin{aligned}
        \Delta\mu^2r^2=z(z+1)\mu^2 r^2/(r_+-r_-)\ll \Delta l(l+1),\\
        \Delta a_*^2\omega^2=z(z+1)\omega^2 a_*^2/(r_+-r_-)\ll \Delta l(l+1).
    \end{aligned}
\end{align}
So in Eq.~(\ref{eq10}), the term
\begin{align}
    \Delta(\mu^2r^2+a_*^2\omega^2+\lambda) \simeq  \Delta\lambda \simeq\Delta l(l+1).
\end{align}
Meanwhile, in the near region, 
\begin{align}
    4{\omega}M r m a_{*} \simeq 4{\omega}M r_+ m a_{*},
\end{align} 
and
\begin{align}
    \omega^2(r^2+a_*^2)^2 \simeq \omega^2(r_+^2+a_*^2)^2=(2{\omega}M r_+)^2,
\end{align}
we get
\begin{align}
    \begin{aligned}
&\omega^2(r^2+a_*^2)^2-4{\omega}Mrm a_{*}+ m^2 a_*^2 
\simeq (ma_*-2{\omega}M r_+ )^2.
    \end{aligned}
\end{align}
Then the radial equation in Eq.~(\ref{eq10}) can be approximated as
\begin{align}\label{eqB16}
    \begin{aligned}
& z(z+1)\frac{d}{dz}\left[z(z+1)\frac{dR}{dz}\right] 
+\big[{\ip}^2-l(l+1)z(z+1)\big]R=0,
    \end{aligned}
\end{align}
where the dimensionless quantity
\begin{align}
{\ip}\equiv\frac{{ma_*-2\omega M r_+}}{{r_+-r_-}}.
\label{pfactor}
\end{align}

In this differential equation (\ref{eqB16}), there are three regular singularities \{\(0, -1, \infty\)\}. The solutions $R(z)$ can be represented by two linear independent solutions \cite{MacRobert1955HigherTF}. Then we can determine the Riemann $P$ form  of the linear independent solutions of Eq.~(\ref{eqB16}) by their index equations
\begin{align}
    x_n(x_n-1)+A_n+B_n=0,
\end{align}
where \(x_n\) is the index of linear independent series solutions. \(A_n\) is the first order characteristic coefficient and \(B_n\) is the second order characteristic coefficient for $n$-th singularity.
Substituting
\begin{align}
    \begin{aligned}
&A_0=1,\qquad B_0={\ip}^2,\\
&A_{-1}=1,\quad~ B_{-1}={\ip}^2,\\
&A_{\infty}=0,\quad~~ B_{\infty}=-l(l+1),
    \end{aligned}
\end{align}
we can get the indexes
\begin{align}
    \begin{aligned}
& x_0^\pm=\pm i {\ip},\quad
x_{-1}^\pm=\pm i {\ip}, \\
& x_{\infty}^+=l+1,\quad 
x_{\infty}^-=- l .
    \end{aligned}
\end{align}
%
The fundamental solutions of $R(z)$ in Eq.~(\ref{eqB16}) can be expressed in the standard form of the  \textit{Riemann P equation} \cite{MacRobert1955HigherTF}: 
\begin{align}
    \begin{aligned}     
R_1(z)&\equiv
    \begin{Bmatrix}
    0   & -1   & \infty & \\
    ip   & -ip   & -l &;z \\
    -ip & ip & l+1& \\
\end{Bmatrix}  \\
&=\left(\frac{z}{z+1}\right)^{i {\ip}}P\begin{Bmatrix}
1 & 0 & \infty& \\
0 & 0 & -l&;z+1 \\
-2i {\ip} & 2i {\ip} & l+1& \\
\end{Bmatrix} \label{eqB22}\\ 
&=
\left(\frac{z}{z+1}\right)^{i {\ip}}F(-l,l+1,1-2i {\ip};z+1),
    \end{aligned}
\end{align}
and $U_1\equiv F(-l,l+1,1-2i {\ip};z+1)$ is the hypergeometry function.
The other corresponding linear independent function is:
\begin{align}
\begin{aligned}
&z^{-2ip}F(-l-2ip,l+1-2ip,1-2i {\ip};z+1)\\&=z^{-2ip}P\begin{Bmatrix}
0   & 1   & \infty & \\
0   & 0   & -l-2ip &;z+1 \\
2ip & 2ip & l+1-2ip& \\
\end{Bmatrix}
.
\end{aligned}
\end{align}
Therefore, the other fundamental solution is
\begin{align}
    \begin{aligned}
    &R_2(z)\\&\equiv \left(\frac{z}{z+1}\right)^{i {\ip}}z^{-2ip}F(-l-2ip,l+1-2ip,1-2i {\ip};z+1),
    \end{aligned}
\end{align}
and we set $U_2\equiv F(-l-2ip,l+1-2ip,1-2i {\ip};z+1)$.

Additionally, we can also get another set of fundamental solutions $(R_3,R_4)$ by transforming the Riemann P form of the hypergeometry function $U_1$, as follows:

\begin{align}
\begin{split}
&\quad F(-l,l+1,1-2i {\ip};z+1)\\
&=P\begin{Bmatrix}
1 & 0 & \infty& \\
0 & 0 & -l&;z+1 \\
-2i {\ip} & 2i {\ip} & l+1& \\
\end{Bmatrix}\\&=(-z^{-1})^{-l}P\begin{Bmatrix}
\infty & 1 & 0& \\
-l & 0 & 0&;-z^{-1} \\
-l-2i {\ip} & 2i {\ip} & 2l+1& \\
\end{Bmatrix}\\&=
(-z)^{l}F(-l,-l-2i {\ip},-2l,-z^{-1})\equiv (-z)^{l} U_3,
\end{split}
\end{align}
and
\begin{align}
\begin{split}
&\quad F(-l,l+1,1-2i {\ip};z+1)\\&=P\begin{Bmatrix}
1 & 0 & \infty& \\
0 & 0 & -l&;z+1 \\
-2i {\ip} & 2i {\ip} & l+1& \\
\end{Bmatrix}\\&=(-z^{-1})^{l+1}P\begin{Bmatrix}
\infty & 1 & 0& \\
l+1 & 0 & -2l-1&;-z^{-1} \\
l+1-2i {\ip} & 2i {\ip} & 0& \\
\end{Bmatrix}\\&=
(-z)^{-l-1}F(l+1,l+1-2i {\ip},2l+2,-z^{-1})\\&\equiv (-z)^{-l-1} U_4.
\end{split}
\end{align}
Thus, the other set of fundamental solutions $(R_3,R_4)$ is
\begin{align}
    \begin{aligned}
        R_3&\equiv\left(\frac{z}{z+1}\right)^{i {\ip}}(-z)^{l}U_3,\\R_4&\equiv\left(\frac{z}{z+1}\right)^{i {\ip}}(-z)^{-l-1} U_4.
    \end{aligned}
\end{align}

Here, we have obtained two sets of fundamental solutions $(R_1,R_2)$ and $(R_3,R_4)$, which correspond to different physical meanings. Specifically, $(R_1,R_2)$ are the ingoing and outgoing solutions near horizon. $(R_3,R_4)$ are the solutions near the centrifugal potential of BHs, which will be used in the following subsection. The full solution $R(z)$ is firstly represented by $(R_3,R_4)$:
 \begin{align}
 \begin{aligned}
R(z)=AR_3+BR_4.
\label{Rr}
 \end{aligned}
 \end{align}
In order to transform Eq.~\eqref{Rr} into the form of $(R_1,R_2)$, we need the relation of the two sets of fundamental solutions. When $z\to0$ (near horizon), we have
\begin{align}
    \begin{aligned}
        R_3=c_1R_1+c_2R_2,\\
        R_4=c_3R_1+c_4R_2,
    \end{aligned}
\end{align}
which are equivalent to 
\begin{align}
    \begin{aligned}
        z^{ip}(-z)^{l}U_3&=c_1U_1(0) z^{ip}+c_2U_2(0) z^{-ip},\\
        z^{ip}(-z)^{-l-1}U_3&=c_3U_1(0)z^{ip}+c_4U_2(0)z^{-ip},
    \end{aligned}
\end{align}
with 
\begin{align}
    \begin{aligned}
        U_1(0)&=F(-l,l+1,1-2i {\ip};1),\\
        U_2(0)&=F(-l-2ip,l+1-2ip,1-2i {\ip};1),
    \end{aligned}
\end{align}
and we determine the coefficients by using the properties of the hypergeometry function
$F(a,b,c;1)=\Gamma(c)\Gamma(c-a-b)/\Gamma(c-a)\Gamma(c-b),F(a,b,c;0)=1$, as follows:
\begin{align}
\begin{aligned}
&c_1=(-1)^{l}\frac{\Gamma(-2l)\Gamma(-2i {\ip})\Gamma(l+1-2ip)\Gamma(-l-2ip)}{\Gamma(-l)\Gamma(-2i {\ip}-l)\Gamma(1-2ip)\Gamma(-2ip)},\\
&c_2=(-1)^{l}\frac{\Gamma(-2l)\Gamma(2i {\ip})\Gamma(l+1)\Gamma(-l)}{\Gamma(-l+2i {\ip})\Gamma(-l)\Gamma(1-2ip)\Gamma(2ip)},\\
&c_3=(-1)^{1-l}\frac{\Gamma(2l+2)\Gamma(-2i {\ip})\Gamma(l+1-2ip)\Gamma(-l-2ip)}{\Gamma(l+1)\Gamma(l-2i {\ip}+1)\Gamma(1-2ip)\Gamma(-2ip)},\\
&c_4=(-1)^{1-l}\frac{\Gamma(2l+2)\Gamma(2i {\ip})\Gamma(l+1)\Gamma(-l)}{\Gamma(l+1)\Gamma(l+2i {\ip}+1)\Gamma(1-2ip)\Gamma(2ip)}.
\end{aligned}
\label{coe}
\end{align}
It should be noted that the above coefficients $c_1,c_3$ and $c_2,c_4$ multiplied by $U_1(0)$ and $U_2(0)$ respectively agree with \cite{Zhou:2023sps}, except for the definition of the sign of $p$.

Finally, the radial solution of Eq.~(\ref{eqB16}) can be expressed as 
\begin{align}
\begin{aligned}
& R(z)=(Ac_1+Bc_3)R_1+(Ac_2+Bc_4)R_2\\&
=(Ac_1+Bc_3)U_1(0)z^{ip}+(Ac_2+Bc_4)U_2(0)z^{-ip}.
\end{aligned}
\end{align}
Using the boundary condition of the event horizon, the reflectivity is
\begin{align}
    \mathcal{R}\equiv\frac{Ac_1+Bc_3}{Ac_2+Bc_4}\frac{U_1(0)}{U_2(0)}.
\end{align}
Then we obtain the relation between \(A\) and \(B\) through the above equation
\begin{align}
    A=B\left(\frac{\mathcal{R}c_4U_2(0)-c_3U_{1}(0)}{c_1U_1(0)-\mathcal{R}c_{2}U_2(0)}\right). \label{ABr}
\end{align}

\subsection{The overlap region solution}

In the overlap region, we connect the solutions in the near region with the ones in the far region to derive the physical solutions. For the property of confluent hypergeometric functions 
\begin{align}
\begin{split}
    &{\hF}(-n-\delta\nu,2l+2;2kr)\\
&=\frac{\Gamma(-2l-1)}{\Gamma(-n-\delta\nu-2n-1)}{\hF}(-n-\delta\nu,2l+2;2kr)\\
&+\frac{(2kr)^{-2l-1}\Gamma(2l+1)}{\Gamma(-n-\delta\nu)}{\hF}(-n-\delta\nu-2l-1,-2l;2kr),
 \end{split}
\end{align}
and Gamma functions
\begin{align}
\begin{aligned}
&\frac{\Gamma(-2l-1)}{\Gamma(-n-\delta\nu-2l-1)}=(-1)^{n}\frac{\Gamma(2l+n+2)}{\Gamma(2l+2)},\\
&\frac{\Gamma(2l+1)}{\Gamma(-n-\delta\nu)}=(-1)^{n+1}\delta\nu \Gamma(n+1)\Gamma(2l+1).
\end{aligned}
\end{align}
So the near asymptotic expansion behavior of the far solutions in the overlap region at \(2kr\to 0\) is
\begin{align}
\begin{split}
    R(r)=&(2kr)^{l}e^{-kr}{\hF}(-n-\delta\nu,2l+2;2kr)
    \\
 \simeq &(-1)^{n}\frac{\Gamma(2l+n+2)}{\Gamma(2l+2)}(2kr)^{l}+\\
 &(-1)^{n+1}\Gamma(n+1)\Gamma(2l+1)\delta\nu(2kr)^{-l-1}.
\end{split}
\label{eqB42}
\end{align}
Meanwhile, when \(r\gg M\) for
\begin{align}
\begin{aligned}
    &R_3=(-z)^{l}U_3 \simeq (-z)^{l}=\left(\frac{-r}{r_+-r_-}\right)^{l},\\
    &R_4=(-z)^{-l-1}U_4 \simeq (-z)^{-l-1}=\left(\frac{-r}{r_+-r_-}\right)^{-l-1},
    \end{aligned}
\end{align}
the remote asymptotic behavior the near solutions in the overlap region is
\begin{align}
    R(r) \simeq  A \left(\frac{-r}{r_+-r_-}\right)^{l}+B \left(\frac{-r}{r_+-r_-}\right)^{-l-1}.
    \label{eqB45}
\end{align}

These two solutions, Eq.~(\ref{eqB42}) and Eq.~(\ref{eqB45}), can be matched to determine $A$ and $B$ in the overlap region. Considering the relation between \(A\) and \(B\) in Eq.~(\ref{ABr}), we can derive the formula of \(\delta\nu\)
\begin{align}\label{eqB46}
    \delta\nu&=\frac{[2k(r_+-r_-)]^{2l+1} \Gamma(-2l)\Gamma(l+1)(\Gamma(2l+n+2)}{\Gamma(-l)\Gamma(2l+1)\Gamma(2l+2)^2 \Gamma(n+1)}\nonumber\\
&\quad \times 
\frac{\Gamma(l-2i {\ip}+1))\Gamma(l+2i {\ip}+1))}{\Gamma(2i {\ip}-l)\Gamma(-l-2i {\ip})}\nonumber\\
&\quad\times \frac{\Gamma(2i {\ip})\Gamma(-l-2i {\ip})-\mathcal{R}\Gamma(-2i {\ip})\Gamma(2i {\ip}-l)}{\Gamma(2i {\ip})\Gamma(l-2i {\ip}+1)-\mathcal{R}\Gamma(-2i {\ip})\Gamma(l+2i {\ip}+1)}\nonumber\\
&=\frac{[2k(r_+-r_-)]^{2l+1} \Gamma(l)\Gamma(l+1)(\Gamma(2l+n+2)}{\Gamma(2l)\Gamma(2l+1)\Gamma(2l+2)^2 \Gamma(n+1)}\nonumber\\
&\quad\times\,2i {\ip} \prod_{j=1}^{l}(j^2+4{\ip}^2)\frac{1-\mathcal{R}\prod_{j=1}^{l}\frac{j-2i {\ip}}{j+2i {\ip}}}{1+\mathcal{R}\prod_{j=1}^{l}\frac{j-2i {\ip}}{j+2i {\ip}}
}.
\end{align}

For the last term, we have
\begin{align}
\begin{aligned}
    \frac{1-\mathcal{R}\prod_{j=1}^{l}\frac{j-2i {p}}{j+2i {p}}}{1+\mathcal{R}\prod_{j=1}^{l}\frac{j-2i {p}}{j+2i {p}}
}
&=\frac{1-|\mathcal{R}|e^{i\phi_{\omega}}}{1+|\mathcal{R}|e^{i\phi_{\omega}}}\\
&=\frac{1-|\mathcal{R}|^2}{1+|\mathcal{R}|^2+2|\mathcal{R}|\cos\phi_{\omega} },
\end{aligned}
\end{align}
where $e^{i\phi_{\omega}}=e^{i\phi}e^{i\text{arg}\mathcal{R}}$ and $e^{i\phi}$ satisfies the relation given by
\begin{align}
    e^{i\phi}=\prod_{j=1}^{l}\frac{j-2i {p}}{j+2i {p}}=\prod_{j=1}^{l}\left(\frac{j-2i {p}}{\sqrt{j^2+4p^2}}\right)^2.
\end{align}
In other words,
\begin{align}
\begin{aligned}
&\phi={i\sum_{j=1}^{l}\ln{\frac{j+2i {p}}{j-2i {p}}}}=2\sum^{l}_{j=1}\phi_{j},
    \\ 
&\phi_{j} \equiv\arctan(-2p/j)\,.
\end{aligned}
\end{align}

\vfill 
\noindent\rule{0.8\textwidth}{0pt} 

\bibliography{Reference} 

\end{document}